\documentclass[12pt]{article}

\usepackage[margin=1in]{geometry}

\usepackage{amssymb}
\usepackage{amsthm}
\usepackage{amsmath}
\usepackage{graphicx}
\usepackage{acronym}
\usepackage{amsfonts}
\usepackage{amscd}

\theoremstyle{plain}
\newtheorem{theorem}{Theorem} [section]
\newtheorem{Example}{Example}
\newtheorem{definition}[theorem]{Definition}
\def\be{\begin{Example}}
\def\ee{\end{Example}}
\def\bt{\begin{theorem}}
\def\et{\end{theorem}\bigskip}
\def\bl{\begin{Lemma}}
\def\el{\end{Lemma}\bigskip}
\def\ep{\end{Proposition}\bigskip}
\def\bp{\begin{Proposition}}
\def\bd{\begin{definition}}
\def\ed{\end{definition}}
\newcommand{\HH}{{\cal H}}
\newcommand{\A}{{\cal A}}

\def\bt{\beta}

\newcommand{\bdes}{\begin{description}}
\newcommand{\edes}{\end{description}}

\newcommand{\bnum}{\begin{enumerate}}
\newcommand{\enum}{\end{enumerate}}

\newcommand{\bit}{\begin{itemize}}
\newcommand{\eit}{\end{itemize}}

\newcommand{\bea}{\begin{eqnarray}}
\newcommand{\eea}{\end{eqnarray}}
\newcommand{\beq}{\begin{equation}}
\newcommand{\eeq}{\end{equation}}

\newcommand{\baray}{\begin{array}}
\newcommand{\earay}{\end{array}}

\newcommand{\bsry}{\begin{subarray}}
\newcommand{\esry}{\end{subarray}}

\newcommand{\bca}{\begin{cases}}
\newcommand{\eca}{\end{cases}}

\newcommand{\bcen}{\begin{center}}
\newcommand{\ecen}{\end{center}}

\newcommand{\bbm}{\begin{bmatrix}}
\newcommand{\ebm}{\end{bmatrix}}

\newcommand{\bmx}{\begin{matrix}}
\newcommand{\emx}{\end{matrix}}

\newcommand{\bpm}{\begin{pmatrix}}
\newcommand{\epm}{\end{pmatrix}}

\newcommand{\btab}{\begin{tabular}}
\newcommand{\etab}{\end{tabular}}

\begin{document}

\title{\bf The Minimum Hartree Value for the Quantum Entanglement Problem}

\author{
Liqun Qi\thanks{Department of Applied Mathematics, The Hong Kong
Polytechnic University, Hung Hom, Kowloon, Hong Kong. E-mail:
\emph{maqilq@polyu.edu.hk}. His work is supported by the Hong Kong
Research Grant Council.}}

\date{\today}
\maketitle

\begin{quote}
{\small \textbf {Abstract.} A general $n$-partite state $| \Psi
\rangle$ of a composite quantum system can be regarded as an element
in a Hilbert tensor product space $\HH = \otimes_{k=1}^n \HH_k$,
where the dimension of $\HH_k$ is $d_k$ for $k = 1, \cdots, n$.
Without loss of generality we may assume that $d_1 \le \cdots \le
d_n$. A separable (Hartree) $n$-partite state $| \phi \rangle$ can
be described by $| \phi \rangle = \otimes_{k=1}^n | \phi^{(k)}
\rangle$ with $| \phi^{(k)} \rangle \in \HH_k$.   We show that
$\sigma := \min \left\{ \langle \Psi | \phi_\Psi \rangle : | \Psi
\rangle \in \HH,\right.$ $\left. \langle \Psi | \Psi \rangle = 1
\right\}$ is a positive number, where $| \phi_\Psi \rangle$ is the
nearest separable state to $| \Psi \rangle$.  We call $\sigma$ the
minimum Hartree value of $\HH$.  We further show that $\sigma \ge
1/{\sqrt{d_1\cdots d_{n-1}}}$. Thus, the geometric measure of the
entanglement content of $\Psi$, $\| | \Psi \rangle - | \phi_\Psi
\rangle \| \le \sqrt{2-2\sigma} \le
\sqrt{2-2\left(1/{\sqrt{d_1\cdots d_{n-1}}}\right)}$. }

\end{quote}

{\large
\section{Introduction}\label{Intr}
\setcounter{equation}{0}

The quantum entanglement problem is now regarded as a central
problem in quantum information processing \cite{HS, NC, WG}.   A
general $n$-partite state $| \Psi \rangle$ of a composite quantum
system can be regarded as an element in a Hilbert tensor product
space $\HH = \otimes_{k=1}^n \HH_k$, where the dimension of $\HH_k$
is $d_k$ for $k = 1, \cdots, n$. Without loss of generality we may
assume that $d_1 \le \cdots \le d_n$.   A natural geometrical
measure of the entanglement content of $| \Psi \rangle$ is the
distance from its nearest separable (Hartree) state \cite{WG}.  The
farther away from the set of separable states, the more entangled a
state is \cite{CXZ, WG}. A separable (Hartree) $n$-partite state $|
\phi \rangle$ can be described by $| \phi \rangle = \otimes_{k=1}^n
| \phi^{(k)} \rangle$ with $| \phi^{(k)} \rangle \in \HH_k$.   In
the next section, we show that $\sigma := \min \left\{ \langle \Psi
| \phi_\Psi \rangle : | \Psi \rangle \in \HH, \langle \Psi | \Psi
\rangle = 1 \right\}$ is a positive number, where $| \phi_\Psi
\rangle$ is the nearest separable state to $| \Psi \rangle$.  We
call $\sigma$ the minimum Hartree value of $\HH$. In Section 3, for
$n=2$, we show that $\sigma = 1/\sqrt{d_1}$.   We further show in
Section 4 that $\sigma \ge 1/{\sqrt{d_1\cdots d_{n-1}}}$ when $n \ge
3$. Thus, the geometric measure of the entanglement content of
$\Psi$, $\| | \Psi \rangle - | \phi_\Psi \rangle \| \le
\sqrt{2-2\sigma} \le \sqrt{2-2\left(1/{\sqrt{d_1\cdots
d_{n-1}}}\right)}$.   Some final remarks are given in Section 5.

\section{The Minimum Hartree Value}
\label{General} \setcounter{equation}{0}

Let $| \Psi \rangle$ be a general $n$-partite pure state of a
composite quantum system.   Then we may denote $| \Psi \rangle \in
\HH$, where $\HH$ is a Hilbert tensor product space $\HH = \HH_1
\otimes \cdots \otimes \HH_n$, and the dimension of $\HH_k$ is $d_k$
for $k = 1, \cdots, n$. We have $\langle \Psi | \Psi \rangle = 1$.
An arbitrary separable $n$-partite state $|\phi \rangle \in \HH$ can
be described by $|\phi \rangle = \otimes_{k=1}^n |\phi^{(k)}
\rangle,$ where $|\phi^{(k)} \rangle \in \HH_k$ and $\| |\phi^{(k)}
\rangle \| = 1$ for $k = 1, \cdots, n$. Denote the set of all
separable states in $\HH$ as $Separ(\HH)$.

For a general $n$-partite state $|\Psi \rangle \in \HH$, a geometric
measure of its entanglement content can be defined as \cite{WG}
\begin{equation} \label{near}
d = \left\| |\Psi \rangle - |\phi_\Psi \rangle \right\| = \min
\left\{ \left\| |\Psi \rangle - |\phi \rangle \right\| : |\phi
\rangle \in Separ(\HH) \right\},
\end{equation}
where $|\phi_\Psi \rangle \in Separ(\HH)$ is the nearest separable
state of $|\Psi \rangle$.  Since the minimization in (\ref{near})
was taken with a continuous function on a compact set in a finite
dimensional space, the nearest separable state to $| \Psi \rangle$
always exists.

For convenience, as in \cite{WG}, instead of studying (\ref{near}),
we may study
\begin{equation} \label{near1}
d^2 = \left\| |\Psi \rangle - |\phi_\Psi \rangle \right\|^2 = \min
\left\{ \left\| |\Psi \rangle - |\phi \rangle \right\|^2 : |\phi
\rangle \in Separ(\HH) \right\}.
\end{equation}
Note that
\begin{equation} \label{more}
\left\| |\Psi \rangle - |\phi \rangle \right\|^2 = \langle \Psi
|\Psi \rangle + \langle \phi |\phi \rangle - \langle \Psi |\phi
\rangle - \langle \phi |\Psi \rangle = 2 - \langle \Psi |\phi
\rangle - \langle \phi |\Psi \rangle.\end{equation} Thus the
minimization problem in (\ref{near1}) is equivalent to the following
maximization problem:
\begin{equation} \label{near2}
\max \left\{ \langle \Psi | \left(\otimes_{k=1}^n |\phi^{(k)}
\rangle \right) + \left( \otimes_{k=1}^n \langle \phi^{(k)} |
\right) | \Psi \rangle : \langle \phi^{(k)} |\phi^{(k)} \rangle = 1,
k=1, \cdots, n \right\}.
\end{equation}

Introducing Lagrange multipliers $\lambda_k$ for $k = 1, \cdots, n$,
we have
\begin{equation} \label{e1}
\langle \Psi | \left(\otimes_{j \not = k}^n |\phi^{(j)} \rangle
\right) = \lambda_k \langle \phi^{(k)} |
\end{equation}
and
\begin{equation} \label{e2}
\left( \otimes_{j \not =k}^n \langle \phi^{(j)} | \right) | \Psi
\rangle= \lambda_k | \phi^{(k)} \rangle.
\end{equation}
We see that
$$\lambda \equiv \lambda_k = \langle \Psi | \phi \rangle = \langle
\phi | \Psi \rangle$$ is a real number in $[-1, 1]$ \cite{WG}.
Then (\ref{e1}) and (\ref{e2}) become
\begin{equation} \label{e3}
\langle \Psi | \left(\otimes_{j \not = k}^n |\phi^{(j)} \rangle
\right) = \lambda \langle \phi^{(k)} |
\end{equation}
and
\begin{equation} \label{e4}
\left(\otimes_{j \not =k}^n \langle \phi^{(j)}| \right) | \Psi
\rangle= \lambda | \phi^{(k)} \rangle.
\end{equation}
Then the largest entanglement eigenvalue $\lambda_*$ corresponds the
nearest separable state $|\phi_\Psi \rangle$, and is equal to the
maximal overlap \cite{WG}:
\begin{equation} \label{near3}
\lambda_* = \langle \Psi | \phi_\Psi \rangle = \max \{ | \langle
\Psi | \phi \rangle | : \phi \in Separ(\HH) \}.
\end{equation}

We now consider the function defined by the maximum function in
(\ref{near3}):
$$g(|z \rangle ) := \max \{ | \langle z |
\phi \rangle | : | \phi \rangle \in Separ(\HH) \},$$ for $| z
\rangle \in \HH$. We see that $g(| z \rangle ) \ge 0$ and $g(| z
\rangle) = 0$ if and only if $| z \rangle  = 0$. Furthermore, for $|
z \rangle, | w \rangle  \in \HH$, we have $g(|z \rangle + | w
\rangle) \le g(| z \rangle ) + g(| w \rangle )$. Hence, $g$ defines
a norm in the finite dimensional space $\HH$. Note that $h(z) =
\sqrt{\langle z | z \rangle}$ also defines a norm in $\HH$.
According to the norm equivalence theorem in the finite dimensional
space \cite{OR}, $\sigma := \min \left\{ \langle \Psi | \phi_\Psi
\rangle \right.$ $ : \left. | \Psi \rangle \in \HH, \langle \Psi |
\Psi \rangle = 1 \right\}$ is a positive number. We call $\sigma$
the minimum Hartree value of $\HH$.   Thus, the geometric measure of
the entanglement content of $|\Psi \rangle$, $\|\Psi \rangle - |
\phi_\Psi \rangle \| \le \sqrt{2-2\sigma}$.

We now summarize this result in the following theorem:

\begin{theorem}
Let the minimum Hartree value of $\HH$ be defined as $\sigma := \min
\left\{ \langle \Psi | \phi_\Psi \rangle : | \Psi \rangle \in \HH,
\langle \Psi | \Psi \rangle = 1 \right\}$, where $| \phi_\Psi
\rangle$ is the nearest separable state to $| \Psi \rangle$.  Then
$\sigma > 0$, and for any $|\Psi \rangle \in \HH$, we have $\langle
\Psi | \phi_\Psi \rangle \ge \sigma$. Furthermore, the geometric
measure of the entanglement content of $| \Psi \rangle$, $\| |  \Psi
\rangle - | \phi_\Psi \rangle \| \le \sqrt{2-2\sigma}$.
\end{theorem}

\section{The Minimum Hartree Value when $n=2$}
\label{two} \setcounter{equation}{0}

We assume that $n=2$.   Let $| e_i \rangle$ for $i = 1, \cdots, d_1$
be an orthonormal basis for $\HH_1$, and $| s_j \rangle$ for $j = 1,
\cdots, d_2$ be an orthonormal basis for $\HH_2$.   Write
$$| \Psi \rangle = \sum_{i=1}^{d_1}\sum_{j=1}^{d_2} \overline{a_{ij}} | e_i
\rangle | s_j \rangle,\ |\phi^{(1)} \rangle = \sum_{i=1}^{d_1} u_i |
e_i \rangle\ {\rm and}\ |\phi^{(2)} \rangle = \sum_{j=1}^{d_2} v_j |
s_j \rangle,$$ where the overbar denotes conjugation.   Then
(\ref{e3}) has the form
$$a_{ij}v_j = \lambda \overline{u_i}$$
and
$$a_{ij}u_i = \lambda \overline{v_j}.$$
Then $A = (a_{ij})$ is a $d_1 \times d_2$ matrix, $u = (u_i)$ is a
$d_1$-dimensional vector and $v = (v_j)$ is a $d_2$-dimensional
vector.   We have \cite{HS}
$$Av = \lambda \overline{u}\ {\rm and}\ A^{\dagger}\overline{u} =
\lambda v,$$ where the dagger denotes the Hermitian conjugate. Then
as in \cite{HS}, we see that $\lambda$ is a singular value of $A$,
and $\lambda_*$ is the largest singular value of $A$.   On the other
hand, since $\langle \Psi | \Psi \rangle = 1$, we see that
$$\sum_{i=1}^{d_1}\sum_{j=1}^{d_2} |a_{ij}|^2 = 1.$$
By linear algebra, $\sigma$, the minimum value of $\lambda_*$ for
all $|\Psi \rangle \in \HH$ with $\langle \Psi | \Psi \rangle = 1$,
is $1 /\sqrt{d_1}$.   Thus, we have the following theorem:
\begin{theorem}
For $n=2$, the minimum Hartree value $\sigma = 1 /\sqrt{d_1}$.
\end{theorem}

Let $a_{jj} = 1/\sqrt{d_1}$ and $a_{ij} = 0$ if $i \not = j$.   Then
we see that $| \Psi \rangle$ is a pure state and $\lambda_* = 1
/\sqrt{d_1} = \sigma$, i.e., the value $\sigma = 1 /\sqrt{d_1}$ is
attainable.

\section{A Lower Bound for the Minimum Hartree Value when $n\ge 3$}
\label{three} \setcounter{equation}{0}

In general, let $| e_{i_k}^{(k)} \rangle$ for $i_k = 1, \cdots, d_k$
be an orthonormal basis for $\HH_k$, $k = 1, \cdots, n$.   Write
$$| \Psi \rangle = \sum_{i_1, \cdots, i_n} \overline{a_{i_1\cdots i_n}} |
e_{i_1}^{(1)} \rangle \cdots | e_{i_n}^{(n)} \rangle,\ {\rm and}\
|\phi^{(k)} \rangle = \sum_{i_k=1}^{d_k} u_{i_k}^{(k)} |
e_{i_k}^{(k)} \rangle.$$ Let $\A$ be a hypermatrix defined by $\A =
\left(a_{i_1\cdots i_n}\right)$.  By (\ref{near3}), we have
\begin{equation} \label{bd}
\lambda_* \equiv \sigma(\A) := \max \left\{ \left| a_{i_1\cdots i_n}
u_{i_1}^{(1)}\cdots u_{i_n}^{(n)} \right| : \| u^{(k)} \|^2 = 1,\
{\rm for}\ k= 1, \cdots, n \right\}.
\end{equation}
Define matrix $A_{i_1\cdots i_{n-2}} = ( A_{i_1\cdots i_{n-2}ij})$
by $A_{i_1\cdots i_{n-2}ij} := a_{i_1\cdots i_{n-2}ij}$.   By
(\ref{bd}),
$$\sigma(A_{i_1\cdots i_{n-2}}) \le \sigma(\A).$$
Let $\| \A \|$ and $\| A_{i_1\cdots i_{n-2}} \|$ be the Frobenius
norm of $\A$ and $A_{i_1\cdots i_{n-2}}$ respectively, i.e.,
$$\| \A \|^2 = \sum_{i_1, \cdots, i_n} | a_{i_1\cdots i_n} |^2,$$
and
$$\| A_{i_1\cdots i_{n-2}} \|^2 = \sum_{i_{n-1},i_n} | a_{i_1\cdots i_n}
|^2.$$ By linear algebra, we have
$$\|A_{i_1\cdots i_{n-2}}\|^2 \le d_{n-1}\sigma(A_{i_1\cdots i_{n-2}}).$$
Since $\langle \Psi | \Psi \rangle = 1$, we have $\| \A \| = 1$.
Putting all of these together, we have
$$1 = \| \A \|^2 = \sum_{i_1,\cdots, i_{n-2}} \| A_{i_1\cdots i_{n-2}}
\|^2 \le \sum_{i_1,\cdots, i_{n-2}} d_{n-1}\sigma(A_{i_1\cdots
i_{n-2}})^2$$
$$\le \sum_{i_1,\cdots, i_{n-2}} d_{n-1}\sigma(\A)^2 = d_1\cdots
d_{n-1} \sigma(\A)^2 = d_1\cdots d_{n-1} \lambda_*^2.$$ By this and
the definition of $\sigma$, we have
$$\sigma \ge 1/\sqrt{d_1\cdots d_{n-1}}.$$
We now have the following theorem:
\begin{theorem}
In general, the minimum Hartree value $\sigma \ge 1/\sqrt{d_1\cdots
d_{n-1}}$.
\end{theorem}
By (\ref{more}), the geometric measure of the entanglement content
of $|\Psi \rangle$, $\|\Psi \rangle - | \phi_\Psi \rangle \| \le
\sqrt{2-2\sigma} \le \sqrt{2-2\left(1/{\sqrt{d_1\cdots
d_{n-1}}}\right)}$.

\section{Final Remarks}
\label{final} \setcounter{equation}{0}

The discussion here follows the spirit of the discussion of the best
rank-one approximation ratio in \cite{Qi}.   The best rank-one
approximation ratio discussion in \cite{Qi} only deals with real
values vectors and hypermatrices.   Also, the minimum Hartree value
here has explicit physical meanings.   These are the differences.
This also stimulates further research to find the exact value of
$\sigma$ when $n \ge 3$.

\bigskip\bigskip

%{\bf Acknowledgment} The author is thankful to Shenglong Hu, Xinzhen Zhang  and two referees for their valuable comments, which
%helped to improve the paper greatly.   Theorem 3.1 is suggested by Shenglong Hu.

\end{document}